\documentclass[submission,copyright,creativecommons]{eptcs}
\providecommand{\event}{PLACES 2016}

\usepackage{breakurl}
\usepackage[linesnumbered,vlined]{algorithm2e}
\usepackage{algorithmic}
\usepackage{amsfonts}
\usepackage{amsmath}
\usepackage{amssymb}
\usepackage{amsthm}
\usepackage[english]{babel} 
\usepackage{csvsimple}
\usepackage{color}
\usepackage{enumitem}
\usepackage{esvect}
\usepackage[ligature, inference]{semantic}
\usepackage{xcolor}

\title{Future-based Static Analysis of Message Passing Programs}
\author{Wytse Oortwijn \qquad\qquad Stefan Blom \qquad\qquad Marieke Huisman
\institute{Formal Methods and Tools, Dept. of EEMCS, University of Twente \\
	P.O.-box 217, 7500 AE Enschede, The Netherlands \\
	\email{\{w.h.m.oortwijn, s.c.c.blom, m.huisman\}@utwente.nl}}
}

\def\titlerunning{Future-based Static Analysis of Message Passing Programs}
\def\authorrunning{W. Oortwijn, S. Blom, and M. Huisman}

\begin{document}
\maketitle

\begin{abstract}
Message passing is widely used in industry to develop programs consisting of several distributed communicating components. Developing functionally correct message passing software is very challenging due to the concurrent nature of message exchanges. Nonetheless, many safety-critical applications rely on the message passing paradigm, including air traffic control systems and emergency services, which makes proving their correctness crucial. We focus on the modular verification of MPI programs by statically verifying concrete Java code. We use separation logic to reason about local correctness and define abstractions of the communication protocol in the process algebra used by mCRL2. We call these abstractions \emph{futures} as they predict how components will interact during program execution. We establish a provable link between futures and program code and analyse the abstract futures via model checking to prove global correctness. Finally, we verify a leader election protocol to demonstrate our approach.
\end{abstract}

\section{Introduction}
\label{sec:intro}

Many industrial applications, including safety-critical ones, consist of several disjoint components that use message passing to communicate according to some protocol. These components are typically highly concurrent, since messages may be sent and received in any order. Developing functionally correct message passing programs is therefore very challenging, which makes proving their correctness crucial~\cite{hoare:2008verified}.

The Message Passing Interface (MPI) is a popular API for implementing message passing programs. We focus on the modular verification of MPI programs, in particular programs written with Java and the MPJ library~\cite{mpj:2016url}. Existing research mainly focusses on proving communication correctness over an abstract system model~\cite{hoare2008:separation,vo:2009formal}. Instead, we target concrete Java source code and combine well-known techniques for static verification with process algebras to reason about communication correctness~\cite{milner:1980calculus}. Communication correctness refers to the correctness of handling messages, i.e. freedom of deadlocks and livelocks, resource leakage, proper matching of sends and receives, etcetera.

Global system correctness depends on the correctness of all individual processes and their interactions. We use permission-based separation logic~\cite{o:2007resources} to statically reason about local correctness. In addition, we model the communication protocol in the mCRL2 process algebra. We refer to these algebraic terms as \emph{futures} as they predict how components will interact during program execution~\cite{zaharieva:2015closer}\footnote{Not to be confused with the concepts of ``futures and promises" used in MultiLisp, described by Halstead~\cite{halstead:1985multilisp} and Liskov and Shrira~\cite{liskov:1988promises}.}. We extend the VerCors toolset~\cite{amighi:2014verification,blom:2014vercors} to establish a provable link between futures and program code so that reasoning about futures corresponds to reasoning about the concrete program. Analysing futures can then be reduced to a parameterised model checking problem to reason about the functional and communication correctness of the system. Since model checkers inherently target finite-state systems, we use abstraction and cut-offs to reason about programs with infinite behaviour whenever possible.

In this paper we analyse the behaviour of several standard MPI operations and model their semantics as process algebra terms. We define several actions, like \textbf{send}, \textbf{recv}, and \textbf{bcast}, that can be used in futures to specify process communication, like that done with Session Types~\cite{honda:2012verification}. After that, we show how to analyse futures, in combination with the process algebra terms that model the semantics of the MPI runtime. The analysis is done via parameterised model checking, for which we plan to use the mCRL2 toolset~\cite{groote:2014modeling}. The key element of our approach is that, by verifying safety and liveness properties on the futures, we actually verify equivalent properties on the concrete system of \emph{any} size. This allows us to check whether process communication always leads to a valid system state in every system configuration. We also plan to check other interesting safety and liveness properties, including: deadlock freedom, resource leakage, and global termination. 

This paper is organised as follows. \autoref{sec:mpi} introduces the message passing paradigm and discusses the semantics of several standard MPI operations. \autoref{sec:comm} contributes an algebraic abstract model of the network environment. \autoref{sec:link} shows how futures and the abstract network environment are used to reason about concrete program code. \autoref{sec:example} demonstrates our approach by verifying a leader election protocol. Finally, \autoref{sec:conclusions} summarises our conclusions and presents future work.

\section{The message passing paradigm}
\label{sec:mpi}

MPI programs consist of a group of interconnected processes $(p_0, \dots, p_{N-1})$ distributed over one or more devices connected via some network. Each process $p_j$ executes the same program in Single Program Multiple Data (SPMD) fashion, in which $j$ is called the \emph{rank} and $N$ the \emph{size}. MPI mainly targets distributed memory systems; processes maintain a private memory and accesses to non-local memories are handled exclusively via message exchanges. Verifying MPI programs involves verifying functional and communication correctness for \emph{every} rank and size. 

The MPI standard describes a set of subroutines to allow processes to exchange messages. We briefly discuss several standard point-to-point and collective MPI operations and their semantics.

\vspace{-8pt}
\paragraph{Standard operations.}
A process $p_i$ can send a message with data element $D$ to process $p_j$ by calling $p_i.\texttt{MPI\_Send}(j, D)$. Similarly, $p_i$ may receive a message from $p_j$ with content $D$ by invoking $D := p_i.\texttt{MPI\_Recv}(j)$. We remove the prefix ``$p_i$." if the sending process can be inferred from the context. Every $\texttt{MPI\_Send}$ must be matched by a $\texttt{MPI\_Recv}$. We allow wildcard receives, i.e. calling $p_i.\texttt{MPI\_Recv}(\star)$ to receive from any process. In this paper we omit tags and communicators.

The MPI standard describes four sending modes for point-to-point communication, namely: blocking mode, buffered mode, synchronous mode, and ready mode. The $p_i.\texttt{MPI\_Send}(j,D)$ operation is a blocking mode send, as it \emph{may} block $p_i$ until matched by a call $p_j.\texttt{MPI\_Recv}(i)$. The buffered variant, $p_i.\texttt{MPI\_BSend}$, does not block $p_i$ until matched by a \texttt{MPI\_Recv} if the request can be buffered. The synchronous variant, $p_i.\texttt{MPI\_SSend}$, \emph{always} blocks $p_i$ until matched by a \texttt{MPI\_Recv}. Finally, the ready mode variant, \texttt{MPI\_RSend}, only succeeds if a matching \texttt{MPI\_Recv} has already been posted. From now, we only consider the blocking mode \texttt{MPI\_Send} variant to illustrate our approach. In future work we also provide support for the other three communication modes.

\vspace{-8pt}
\paragraph{Non-blocking variants.}
The immediate send, \texttt{MPI\_ISend}, is an immediate variant on the standard blocking send. Calling $r := p_i.\texttt{MPI\_ISend}(j,D)$ is \emph{not} allowed to block $p_i$ for a matching \texttt{MPI\_Recv} by $p_j$. Instead, \texttt{MPI\_ISend} returns a request handle $r$ which can be used to query or influence the status of the request. The operation $p_i.\texttt{MPI\_Wait}(r)$ blocks $p_i$ until the operation corresponding to the handle $r$ has been completed. Two consecutive non-blocking calls $p_i.\texttt{MPI\_ISend}(j, \star)$ and $p_i.\texttt{MPI\_ISend}(k, \star)$ made by process $p_i$ are \emph{forced} to match in order if $j=k$, but may be handled in \emph{any} order if $j \not = k$. 

Apart from the immediate blocking send, also the other three sending modes have immediate variants, which are: \texttt{MPI\_IBsend}, \texttt{MPI\_ISsend}, and \texttt{MPI\_IRsend}. In this paper we only focus on \texttt{MPI\_ISend}. In future work we provide support for the other three immediate operations.

\vspace{-8pt}
\paragraph{Collective operations.}
The collective $p_i.\texttt{MPI\_Barrier}()$ operation blocks execution of $p_i$ until matched by a call $p_j.\texttt{MPI\_Barrier}()$ for every $j \not = i$ and can therefore be used to synchronise all processes. Non-blocking operations invoked before calling $\texttt{MPI\_Barrier}$ may complete while lingering in the barrier, as they are handled by the network environment. We plan to provide support for other collective operations, like: $\texttt{MPI\_Reduce}$, $\texttt{MPI\_Scatter}$, and $\texttt{MPI\_Gather}$, but this is future work. The collective $p_i.\texttt{MPI\_Bcast}(D)$ operation is used to broadcast a data element $D$ to every participating process. For now we assume that $p_i.\texttt{MPI\_Bcast}(D)$ simply calls $p_i.\texttt{MPI\_Send}(j, D)$ for every $j \not = i$.

\section{Modelling the communication network}
\label{sec:comm}

We use futures to predict the communication protocol of MPI programs. In this section we specify abstract actions, e.g. $\textbf{send}$, $\textbf{recv}$, and $\textbf{bcast}$, that correspond to concrete MPI operations. We also model the network environment and thereby the semantic behaviour of the MPI operations. Futures are constructed via the following syntax, which is a subset of the multi-action process algebra used in mCRL2, proposed by Groote and Mousavi~\cite{groote:2014modeling}:

\vspace{-2pt}
$$P ::= \textbf{a} ~|~P + P ~|~ P \cdot P ~|~ P \parallel P ~|~ c \rightarrow P ~|~ c \rightarrow P \diamond P ~|~ \sum_{d \in T} P(d) ~|~ \textsf{X}(u_1, \dots, u_n)$$

\vspace{-1pt}
Actions have the form $\textbf{a} : T_1 \times \cdots \times T_n$, where all $T_i$ are types, e.g. sets, maps, integers, sequences, and so on. We include $\tau$ as a special silent action. Sequential composition is denoted by $P \cdot P'$ ($P$ is executed before $P'$) and choice is denoted by $P + P'$ (either $P$ or $P'$ is executed). Summations $\sum_{d\in T}P(d)$ of choices are also supported, where $P$ is executed for some model $d$ of type $T$. The expression $P \parallel P'$ means putting $P$ and $P'$ in parallel, thereby allowing their actions to be interleaved. Conditions are written either as $c \rightarrow P$ (execute $P$ only if $c$ is true) or $c \rightarrow P \diamond P'$ (execute $P$ if $c$ is true, otherwise execute $P'$). Finally, $\textsf{X}(u_1, \dots, u_n)$ calls a process $\textsf{X}$ with a list of $n$ arguments of appropriate types.

Two actions $\textbf{a},\textbf{b}$ may communicate by specifying a \emph{communication pair}, written $\textbf{a}|\textbf{b}$. Two communicating actions are forced to happen simultaneously and thereby impose restrictions on the evaluation of actions. As an effect, communication pairs introduce synchronisation points between processes.

\vspace{-4pt}
\paragraph{Modelling the network environment.}
Recall that MPI programs are executed on a group of $N$ processes $(p_0, \dots, p_{N-1})$. Let $\mathcal{R} = \{0, \dots, N - 1\}$ be the set of ranks. The network can be modelled as a collection of queues that store pending messages, i.e. messages sent but not yet received. Therefore, we model the network environment as a recursive process, named \textsf{Network}, that maintains a set $T = \{ Q_{i,j} \ | \ i,j \in \mathcal{R} \}$ of queues $Q_{i,j}$ for each pair of ranks $i,j \in \mathcal{R}$ to maintain the state of the network. \textsf{Network} is defined as follows, where $\mathcal{M}$ denotes the set of all possible messages:

\vspace{-8pt}
\begin{equation*}
	\begin{split}
		\textbf{process} \ \textsf{Network}(T) \equiv \sum_{i,j\in\mathcal{R}} & \sum_{m \in \mathcal{M}} \Big( \textbf{nrecv}(i,j,m) \cdot \textsf{Network}(T.enqueue(i, j, m)) \ + \ \\
		&T.peek(i, j, m) \rightarrow \textbf{nsend}(j,i,m) \cdot \textsf{Network}(T.dequeue(i, j)) \Big)
	\end{split}
\end{equation*}

Observe that \textsf{Network} uses two actions, namely: (1) \textbf{nsend} for sending a message from the network to a process; and (2) \textbf{nrecv} for receiving a message from a process. Processes may use the network by communicating with these two actions. More specifically, we define two actions, \textbf{send} and \textbf{recv}, that correspond to the functions \texttt{MPI\_Send} and \texttt{MPI\_Recv}, respectively. Standard blocking-mode sends and receives are performed via the communication pairs $\textbf{send}|\textbf{nrecv}$, $\textbf{recv}|\textbf{nsend}$, and $\textbf{send}|\textbf{recv}$. For the immediate blocking send, \texttt{MPI\_ISend}, we also define a corresponding action $\textbf{isend}$ and use the communication pair $\textbf{isend}|\textbf{nrecv}$. Moreover, the $T.enqueue(i,j,m)$ and $T.dequeue(i,j)$ functions are just auxiliary functions used to enqueue/dequeue elements from $Q_{i,j}$, which can easily be specified in mCRL2. In the remaining of this paragraph, we give more detail on the abstract actions and communication pairs.

When a future performs the $\textbf{send}(i,j,m)$ action, which corresponds to a call $p_i.\texttt{MPI\_Send}(j, m)$ in the program code, the network may receive the message via communication with $\textbf{nrecv}(i,j,m)$, i.e. by having the communication pair $\textbf{send}|\textbf{nrecv}$. After communication, the message is stored in the network by applying $T.enqueue(i,j,m)$, which enqueues $m$ onto $Q_{i,j}$. Similarly, the $\textbf{recv}(i,j,m)$ action, which corresponds to the invocation $m := p_i.\texttt{MPI\_Recv}(j)$, communicates with $\textbf{nsend}$, i.e. $\textbf{recv}|\textbf{nsend}$. The network environment can send a message $m$ to process $p_i$ via the $\textbf{nsend}(i,j,m)$ action if $p_i$ chooses to communicate with the network via a matching $\textbf{recv}(i,j,m)$ action. The network can only send the top element of $Q_{i,j}$ as the MPI standard enforces a FIFO order, hence the check $T.peek(i,j,m)$, which returns true only if $m$ is the top element of $Q_{i,j}$. When $p_i$ chooses to receive the message $m$, the queue $Q_{i,j}$ is updated by dequeuing the message $m$, which is done by applying $T.dequeue(i,j,m)$. Since $p_i.\texttt{MPI\_Send}$ \emph{may} block $p_i$ until matched by a \texttt{MPI\_Recv} but is not forced to do so, we also allow \textbf{send} to communicate directly with \textbf{recv}, i.e. having the communication pair $\textbf{send}|\textbf{recv}$, thereby bypassing the network process.

For the immediate blocking send, \texttt{MPI\_ISend}, we define a corresponding action $\textbf{isend}$, so that calling $r := p_i.\texttt{MPI\_ISend}(j, D)$ corresponds to $\textbf{isend}(i,j,m,r)$. The $\textbf{isend}$ action can not directly communicate with $\textbf{recv}$ and therefore \emph{only} communicates via the network: $\textbf{isend}|\textbf{nrecv}$. Note that two consecutive actions $\textbf{isend}(i, j, m, r)$ and $\textbf{isend}(i, k, m', r')$ performed by rank $i$ match in order if $j=k$, since $m$ and $m'$ are consecutively added to the same queue $Q_{i,j}=Q_{i,k}$ and thus handled in that specific order. The two actions may perform in any order if $j\not=k$, as they are added to different queues $Q_{i,j} \not = Q_{i, k}$. 

To support other send modes, the \textsf{Network} process is still too simple. For example, to support the buffered sends \texttt{MPI\_Bsend} and \texttt{MPI\_IBsend}, buffer space needs to be taken into account, which may potentially break the FIFO ordering. Ready-mode sends require extra checks to ensure that the network already contains a matching receive. Synchronous-mode sends only allow direct synchronisations, e.g. having the communication pair $\textbf{ssend}|\textbf{recv}$, where \textbf{ssend} corresponds to \texttt{MPI\_Ssend}.

\vspace{-4pt}
\paragraph{Short example of network interaction.}
To illustrate the use of multi-actions to communicate with the network environment, we give a short producer/consumer example. Suppose that the network is used by two processes: (1) a producer that only sends messages; and (2) a consumer that only receives messages sent by the producer. The producer and consumer are defined as follows:

$$\textbf{process}~\textsf{Producer}(v:\mathbb{Z}) \equiv \textbf{send}(0, 1, v) \cdot \textsf{Producer}(v + 1)$$
$$\textbf{process}~\textsf{Consumer}(t:\mathbb{Z}) \equiv \sum_{n \in \mathbb{Z}} \textbf{recv}(1, 0, n) \cdot \textsf{Consumer}(t + n)$$

Consider the initial configuration $\textsf{Producer}(0) \parallel \textsf{Consumer}(0) \parallel \textsf{Network}(T_{\emptyset})$, where $T_{\emptyset}$ denotes the set of empty queues, i.e. $|Q_{i,j}| = 0$ for every $Q_{i,j} \in T_{\emptyset}$. From the initial configuration, the only allowed transition is the multi-action $\textbf{send}|\textbf{nrecv}$, which results into the configuration $\textsf{Producer}(1) \parallel \textsf{Consumer}(0) \parallel \textsf{Network}(T')$, where $T'$ is equal to $T_{\emptyset}$, but with $0$ enqueued onto $Q_{0,1}$. From this configuration, either $\textbf{send}|\textbf{nrecv}$ (the producer sends another value to the network) or $\textbf{recv}|\textbf{nsend}$ (the consumer receives a value from the network) may happen, and this process repeats forever.

\vspace{-4pt}
\paragraph{Modelling broadcasts and barriers.}
For broadcasting we define the action $\textbf{bcast}(i,m)$ that corresponds to the function call $p_i.\texttt{MPI\_Bcast}(m)$. Broadcasts are handled by a separate process, called $\textsf{Bcast}$, dedicated to transform a call $p_i.\texttt{MPI\_Bcast}(m)$ into a series of calls $p_i.\texttt{MPI\_Send}(j, m)$ for every $j \not = i$. The $\textsf{Bcast}$ process is defined as follows:

\vspace{-8pt}
\begin{equation*}
	\begin{split}
		\textbf{process} \ \textsf{Bcast}() &\equiv \sum_{i \in \mathcal{R}} \sum_{m \in \mathcal{M}} \Big( \textbf{breq}(i, m) \cdot \textsf{Handle}(i,m,\mathcal{R}\backslash \{i\}) \Big) \\
		\textbf{process} \ \textsf{Handle}(i,m,R) &\equiv (R \not = \emptyset) \rightarrow \sum_{j \in R} \Big( \textbf{nsend}(i,j,m) \cdot \textsf{Handle}(i,m,R \backslash \{j\}) \Big) \diamond \textsf{Bcast}()
	\end{split}
\end{equation*}

When a future starts broadcasting by performing $\textbf{bcast}(i,m)$, the \textsf{Bcast} process receives the broadcast request by communicating with $\textbf{breq}$, i.e. via the communication pair $\textbf{bcast}|\textbf{breq}$. The \textsf{Handle} process actually handles the broadcast request by generating a $\textbf{nsend}(i,j,m)$ action for every $j \not = i$. Any process $p_j$ may then receive the message $m$ by communicating via a standard \textbf{recv} action. Note that $\textsf{Handle}(i,m,R)$ only calls $\textsf{Bcast}$ if all ranks $j \in \mathcal{R} \backslash \{i\}$ have synchronised on $\textbf{nsend}(i,j,m)$. 

For handling barriers we define the action $\textbf{barrier}(i)$ that corresponds to the call $p_i.\texttt{MPI\_Barrier}()$. Similar to broadcasts, also barriers are handled by a dedicated process, named \textsf{Barrier}, which simply synchronises with the $\textbf{barrier}$ action. In particular, performing an action $\textbf{barrier}(i)$ prevents further actions to be executed until $\textsf{Barrier}$ has synchronised with $\textbf{barrier}(j)$ for every $j \not = i$. We omit the definition of \textsf{Barrier}, as it is essentially the same as \textsf{Bcast}. 

\section{Linking futures to program code}
\label{sec:link}

We use permission-based separation logic~\cite{o:2007resources} to reason about local correctness of MPI programs. We include the permissions, since we allow MPI processes to create threads. In particular, we assume that MPI programs $S$ have preconditions $P$ and postconditions $Q$, so that partial correctness can be proven via Hoare triples $\{P\} S \{Q\}$. The conditions $P$ and $Q$ may then use permissions to guarantee data-race freedom in case of multi-threading.

\vspace{-4pt}
\paragraph{Hoare triple reasoning.}
To find a correspondence between program code and futures, we use Hoare triple reasoning. More specifically, we extend Hoare triples to verify that an MPI program satisfies its algebraically predicted behaviour. For example, we may specify a future: ``$\textbf{send}(i,j,m) \cdot \textbf{recv}(j,i,n)$'' that predicts the behaviour of the program fragment: ``$p_i.\texttt{MPI\_Send}(j, m); ~n := p_i.\texttt{MPI\_Recv}(j)$''. We prove a link between the future and the program fragment via the Hoare triples: $\{ \textbf{send}(i,j,m) \cdot \textbf{recv}(j,i,n) \cdot \textsf{F} \} p_i.\texttt{MPI\_Send}(j, m) \{ \textbf{recv}(j,i,n) \cdot \textsf{F} \}$ and $\{ \textbf{recv}(j,i,n) \cdot \textsf{F} \} n = \texttt{MPI\_Recv}(j) \{ \textsf{F} \}$, where $\textsf{F}$ describes the future of the remaining program. To generalise, for the \textbf{send}, \textbf{recv}, \textbf{bcast}, and \textbf{barrier} actions we use the following four Hoare triple axioms: \\

\noindent
{
\small
\begin{tabular}{ll}
	$$\inference[\text{[send]}~:~]{}{\{ \textbf{send}(i,j,m) \cdot \textsf{F} \} p_i.\texttt{MPI\_Send}(j,m) \{ \textsf{F} \}}$$ & $$\inference[\text{[recv]}~:~]{}{\{ \textbf{recv}(i,j,m) \cdot \textsf{F} \} m := p_i.\texttt{MPI\_Recv}(j) \{ \textsf{F} \}}$$ \\
\end{tabular}
} \\

\noindent
{
\small
\begin{tabular}{ll}
	$$\inference[\text{[bcast]}~:~]{}{\{ \textbf{bcast}(i,m) \cdot \textsf{F} \} p_i.\texttt{MPI\_Bcast}(m) \{ \textsf{F} \}}$$ & $$\inference[\text{[barrier]}~:~]{}{\{ \textbf{barrier}(i) \cdot \textsf{F} \} p_i.\texttt{MPI\_Barrier}() \{ \textsf{F} \}}$$ \\
\end{tabular}
}

\vspace{12pt}
All other actions that correspond to MPI functions, e.g. immediate sends like \textbf{isend}, alternative send modes like \textbf{ssend}, \textbf{bsend}, etcetera, are proven to correspond to the MPI program in the same way. We extend the VerCors toolset to verify these Hoare triples by specifying appropriate triples to handle sequential composition of futures: $\textsf{F} \cdot \textsf{G}$, choices between futures: $\textsf{F} + \textsf{G}$, and so on. 

\paragraph{Splitting futures.}
Since we allow multi-threading, we use permission-based separation logic to prove local correctness of MPI programs. The process algebraic equivalent to multi-threading is the parallel composition: $\textsf{F} \parallel \textsf{G}$. We assign permissions to futures, written ``$\text{Future}(\pi, \textsf{F})$'', meaning that a permission fragment $\pi$ is assigned to the future $\textsf{F}$. Then we allow splitting and merging of futures: ``$\text{Future}(\pi_1 + \pi_2, \textsf{F} \parallel \textsf{G}) *-* \text{Future}(\pi_1, \textsf{F}) * \text{Future}(\pi_2, \textsf{G})$'' in separation logic style. The splitted futures can then be distributed among parallel threads, where the permissions can be used for allocating resource invariants.

We refer to the initial future as the \emph{global} future, since it has full permission. The global future can be split into \emph{local} futures with fractional permissions. Proving correctness of local futures with respect to a program fragment is done via standard Hoare logic reasoning~\cite{zaharieva:2015closer}. 
For example, the algorithm in \autoref{fig:protocol}, which is discussed in detail in \autoref{sec:example}, shows how program code is annotated with futures. In \autoref{fig:protocol}, all futures have full permission since the algorithm does not fork additional threads.

\vspace{-4pt}
\paragraph{Communication correctness.}
Let $P$ be an MPI program and $\textsf{F}$ its predicted global future. If we can prove that $P$ correctly executes according to $\textsf{F}$ (i.e. proving $\{ \textsf{F} \}P \{ \epsilon \}$ for the empty process $\epsilon$), then we need to analyse $\textsf{F} \parallel \cdots \parallel \textsf{F}$ to reason about functional and communication correctness of the network of processes all running $P$. Let $\textsf{F}^n = \textsf{F} \parallel \cdots \parallel \textsf{F}$ be the parallel composition of $n$ futures $\textsf{F}$. In particular, we need to verify correctness of $\textsf{F}^N$ for \emph{every} size $N$, in combination with the network environment. Therefore, we will use mCRL2~\cite{mcrl2:2016url} to analyse the following configuration, where $T_{\emptyset}$ denotes the set of empty queues: 
$$\textsf{F}^N \parallel \Big( \textsf{Network}(T_{\emptyset}) \parallel \textsf{Bcast}() \parallel \textsf{Barrier}() \Big)$$

\section{Example: A leader election protocol}
\label{sec:example}

We illustrate our approach on a small example with $N$ processes $(p_0, \dots, p_{N-1})$ performing a leader election protocol. The processes communicate in a ring topology, so that process $p_i$ only sends messages to $p_{i + 1}$ and only receives messages from $p_{i-1}$, counting modulo $N$. Each process $p_i$ holds a \emph{unique} integer $v_i$ so that $v_i \not = v_j$ for $i \not = j$. The leader is the process $p_j$ with the highest value $v_j$, that is, $v_j = max \{ v_0, \dots, v_{N-1} \}$. The protocol operates in a number of rounds. In each round, each process circulates and remembers its highest encountered value. Ultimately, after $N$ rounds, all processes know the highest participating value $v_j$ and the leader announces itself by broadcasting its rank.

\begin{figure}[t]
	\centering
		\begin{tabular}{cc}
		\begin{minipage}{.5\textwidth}
			\begin{algorithm}[H]
				\SetKwProg{Fn}{def}{\string:}{}
				\SetKwFunction{fun}{election}
				\SetKw{KwTo}{in}\SetKwFor{For}{for}{\string:}{}
				\SetKwIF{If}{ElseIf}{Else}{if}{}{elif}{else}{}
				\SetKwFor{While}{while}{:}{fintq}
				\SetInd{0.5em}{0.5em}
				\AlgoDontDisplayBlockMarkers
				\SetAlgoNoEnd
				\SetAlgoNoLine

				{\color{purple} \textbf{requires} $0 \leq n \leq N~\wedge~h \geq v$} \\
				{\color{purple} \textbf{requires} Future$(\textsf{Elect}(i, h, v, n) \cdot \textsf{F})$} \\
				{\color{purple} \textbf{ensures} Future$(\textsf{F})$} \\
				\Fn{\fun{\textup{\textbf{int}} $h$, \textup{\textbf{int}} $v$, \textup{\textbf{int}} $n$}} {
					$\textbf{int}~i \gets \texttt{MPI\_Rank}()$ \\
					$\textbf{int}~N \gets \texttt{MPI\_Size}()$ \\
					$\texttt{MPI\_Send}(i + 1~\text{mod}~N,~{\color{blue}\texttt{elect} \langle h \rangle})$ \label{line:sendelect} \\
					${\color{blue}\texttt{elect} \langle h' \rangle} \gets \texttt{MPI\_Recv}(i - 1~\text{mod}~N)$ \label{line:recvelect} \\
					\textbf{if} $h < h'$ \textbf{then} $h \gets h'$ \label{line:updateh} \\
					\textbf{if} $n < N$ \textbf{then} $\texttt{election}(h,~v,~n+1)$ \label{line:reccheck} \\
					\textbf{else} $\texttt{chooseLeader}(h,~v)$ \\
				}
			\end{algorithm}
		\end{minipage}
		&
		\begin{minipage}{.5\textwidth}
			\begin{algorithm}[H]
				\SetKwProg{Fn}{def}{\string:}{}
				\SetKwFunction{fun}{chooseLeader}
				\SetKw{KwTo}{in}\SetKwFor{For}{for}{\string:}{}
				\SetKwIF{If}{ElseIf}{Else}{if}{}{elif}{else}{}
				\SetKwFor{While}{while}{:}{fintq}
				\SetInd{0.5em}{0.5em}
				\AlgoDontDisplayBlockMarkers
				\SetAlgoNoEnd
				\SetAlgoNoLine

				{\color{purple} \textbf{requires} $h \geq v$} \\ 
				{\color{purple} \textbf{requires} Future$(\textsf{Choose}(i, h, v) \cdot \textsf{F})$} \\
				{\color{purple} \textbf{ensures} Future$(\textsf{F})$} \\
				\Fn{\fun{\textup{\textbf{int}} $h$, \textup{\textbf{int}} $v$}} {
					$\textbf{int}~i \gets \texttt{MPI\_Rank}()$ \\
					\textbf{if} $h = v$ \textbf{then} \\
					\Indp $\texttt{MPI\_Bcast}({\color{blue}\texttt{lead} \langle i \rangle})$ \label{line:bcastrank} \\
					\Indm \textbf{else} \\
					\Indp ${\color{blue} \texttt{lead} \langle j \rangle} \gets \texttt{MPI\_Recv}(\star)$ \label{line:recvrank} \\
					\Indm $\texttt{MPI\_Barrier()}$ \\
				}
			\end{algorithm}
		\end{minipage}
	\end{tabular}
	\vspace{-8pt}
	\caption{Annotated example program of a leader election protocol with simplified MPI syntax.}
	\vspace{-2pt}
	\label{fig:protocol}
\end{figure}

\autoref{fig:protocol} shows annotated pseudocode. The program distinguishes between two kinds of messages: $\color{blue}\texttt{elect}\langle n \rangle$ for communicating an integer $n \in \mathbb{Z}$, and $\color{blue}\texttt{lead}\langle j \rangle$ for communicating a rank $j \in \mathcal{R}$. Below the predicted futures $\textsf{Elect}$ and $\textsf{Choose}$ are given for $\texttt{election}$ and $\texttt{chooseLeader}$, respectively.

\vspace{-4pt}
\begin{equation*}
	\begin{split}
		\textbf{process} & \ \textsf{Elect}(i,v,h,n) \equiv (n < N) \rightarrow  \sum_{h' \in \mathcal{M}} \Big( \textbf{send}(i,i+1 ~\text{mod}~ N,{\color{blue}\texttt{elect}\langle h \rangle}) \\ & \cdot \textbf{recv}(i - 1 ~\text{mod}~ N, i, {\color{blue}\texttt{elect}\langle h' \rangle}) \cdot \textsf{Elect}(i, v, max(h,h'), n + 1) \Big)  \diamond \textsf{Choose}(i,h,v) \\
		\textbf{process} & \ \textsf{Choose}(i,h,v) \equiv \Big( (h=v) \rightarrow \textbf{bcast}(i, {\color{blue} \texttt{lead} \langle i \rangle}) ~\diamond~ \sum_{j \in \mathcal{R}} \textbf{recv}(j,i,{\color{blue}\texttt{lead} \langle j \rangle}) \Big) \cdot \textbf{barrier}(i)
	\end{split}
\end{equation*}

Assume that each process $p_i$ is started by invoking $\texttt{election}(v_i, v_i, 0)$ and therefore starts with the initial future $\textsf{Elect}(i, v_i, v_i, 0)$. We \emph{predict} with \textsf{Elect} that $p_i$ sends the value $h$ to process $p_{i+1}$ in an $\color{blue}\texttt{elect}$ message, receives a value $h'$ from process $p_{i-1}$ as an $\color{blue}\texttt{elect}$ message, and repeats this process as long as $n < N$ while considering $h \gets max(h, h')$. After $N$ rounds we predict the leader $p_j$ to broadcast its rank by sending a $\color{blue}\texttt{lead}\langle j \rangle$ message and all other processes $p_i$ to receive the $\color{blue}\texttt{lead} \langle j \rangle$ message.

By observing the code, a leader is elected by circulating all values $v_i$ through the ring topology in $N$ rounds, hence the check at \autoref{line:reccheck}. In each round, every process $p_i$ receives a message $\color{blue}\texttt{elect}\langle h' \rangle$ from $p_{i-1}$ (\autoref{line:recvelect}), where $h'$ is the maximum value encountered by $p_{i-1}$. In turn, $p_i$ send its highest encountered value $h$ to $p_{i+1}$ (\autoref{line:sendelect}). Finally, $h$ is updated (\autoref{line:updateh}) to remain the highest encountered value before starting the next round (\autoref{line:reccheck}). After $N$ rounds, each process $p_i$ invokes the function $\texttt{chooseLeader}(h,v)$. If $h = v$, then $p_i$ is the leader due to the uniqueness of the values $v_i$. The leader $p_j$ broadcasts its rank $j$ to all other processes via a $\color{blue}\texttt{lead} \langle j \rangle$ message (\autoref{line:bcastrank}). All other processes receive the rank of the leader (\autoref{line:recvrank}). Finally, all processes synchronise by entering a barrier. 

The $\texttt{election}$ function is proven via the following Hoare triple: $\{ 0 \leq n \leq N \ \wedge \ v \leq h \ \wedge \ \textsf{Elect}(i,v,h,n) \cdot \textsf{F} \} p_i.\texttt{election}(h,v,n) \{ \textsf{F} \}$, with $p_i$ the executing process. Similarly, the $\texttt{chooseLeader}$ function is proven via the triple: $\{ h \geq v \ \wedge \ \textsf{Choose}(i,h,v) \cdot \textsf{F} \} p_i.\texttt{chooseLeader}(h,v) \{ \textsf{F} \}$. After that, we use mCRL2 to reason about the global state and thereby reason about program executions in a global setting. For example, we may verify that the invocation $\texttt{chooseLeader}(h,v_i)$ receives a parameter $h \in \mathbb{Z}$ such that $h = max \{v_0, \dots, v_{N-1} \}$ and $h = v_j$ for exactly one $j \in \mathcal{R}$. In particular, we verify that the leader $p_j$ eventually broadcasts the message $\color{blue}\texttt{lead}\langle j \rangle$ containing its rank, and that all other processes eventually receive that message. In that case, all processes receive the rank of the leader and communicate correctly according to the predicated future.

\section{Conclusion and future work}
\label{sec:conclusions}

The work described in this paper is still in its early stages. We have already manually worked out a couple of verification examples, including the leader election protocol described in \autoref{sec:example}. We are currently working on providing tool support by extending the VerCors toolset. In particular, we define new Hoare triples to handle futures in combination with the abstract MPI actions described in \autoref{sec:comm}. Moreover, we are extending the network environment to support more MPI functions, like: \texttt{MPI\_Scatter}, \texttt{MPI\_Gather}, and \texttt{MPI\_Reduce}, which are focused on distributing data in a specific way. After extending VerCors we plan to make a connection with mCRL2 for future-based analysis of the global system consisting of $N$ instances of the program. Finally, we plan to show correctness of some industrial message passing programs in a case study. \\

\noindent
{\large \textbf{Acknowledgements.}} Oortwijn is funded by the NWO TOP project VerDi (projectnr. 612.001.403). Blom and Huisman are funded by the ERC 258405 VerCors project.

\nocite{*}
\bibliographystyle{eptcs}
\vspace{-8pt}
\bibliography{references}

\end{document}